\documentclass[12pt]{article}
\usepackage{epsfig}
\usepackage{amsmath}

\begin{document}
\title
{\bf Zero width resonance (spectral singularity) in a complex PT-symmetric potential}
\author{Zafar Ahmed\\
Nuclear Physics Division, Bhabha Atomic Research Centre,
Bombay 400 085\\
zahmed@barc.gov.in}
\date{\today}
\maketitle
\begin{abstract}
We show that the complex PT-Symmetric potential, $V(x)=-V_1 \mbox{sech}^2x ~+~ iV_2~ \mbox{sech}x 
~\tanh x, $, entails a single zero-width resonance (spectral singularity) when 
$V_1+|V_2|=4n^2+4n+{3\over 4}~(n=1,2,3.., |V_2|>|V_1|+ {\mbox{sgn}(V_1) \over 4})$ and the positive resonant 
energy is given as $E_*={1 \over 4}[|V_2|-(1/4+V_1)]$.
\\ \\
PACS No.: 03.65.Ge, 03.65.Nk, 11.30.Er, 42.25.Bs
\end{abstract}
The physical energy poles of s-matrix or transmission/reflection amplitudes yield
discrete spectrum of bound states and resonances of a Hermitian scattering potential well.
In the former, the energies are real and negative (within the potential well),
in the latter these are complex where the real part is positive. These are
are also called Gamow or Sigert states embedded in positive energy continuum.
For about a decade [1,2] now, the non-Hermitian complex PT-symmetric potentials
have been investigated to have a real discrete spectrum. The PT-symmetric potentials
are invariant under joint action of P(arity: x$\rightarrow$ -x) and 
T(ime-reversal: i $\rightarrow $-i).
\par The complex PT-symmetric potential
\begin{equation}
V(x)=-V_1 \mbox{sech}^2 x + i V_2 \mbox{sech} x \tanh x.
\end{equation}
is the first [3] exactly solvable model of complex PT-symmetric potential to
analytically and explicitly demonstrate that the spectrum is real and discrete
provided $V_2 < V_1+1/4$ (assuming $V_1$ to be positive) and the energy eigenstates
are also the eigenstates of the antilinear
operator PT, other wise the PT-symmetry is spontaneously broken and the spectrum contains
complex conjugate pairs of eigenvalues. This model has helped in finding or demonstrating several
other features of complex PT-symmetric interactions [4]
\par The real Hermitian version of this scattering potential is called Scarf II for which
the exact analytic scattering amplitudes have already been found [4,5]. In this Letter, we
would like to show that the reflection/transmission amplitudes for (1) when $V_1>0$ have two kinds of
discrete poles one set of them are having real and complex-conjugate energies with real
part as negative. These are othewise known as discrete spectrum of bound states [3]. 
\par The other one is a single positive energy which exists provided the potential
parameters $V_1,V_2$ satisfy a certain special condition. This is quite like the shape
resonance embedded in positive energy continuum of a Hermitian potential. However, contrastingly
the new resonance is having a zero width. In recent instructive investigations [7,8], this pole is
discussed as a spectral singularity of non-Hermitian Hamiltonian which is also like a resonance with 
zero width. In Ref. [8], as an example a complex PT-symmetric
model has been used to find the spectral singularity, the calculations are very cumbersome and implicit.
In the following, we present the potential (1) as an exactly solvable model for the spectral 
singularity. Here both the condition on the potential parameters and the resonant energy are
very simple and explicit.
\par Using $2m=1=\hbar^2$ for the Schr{\"o}dinger equation let us define
\begin{equation}
r={1\over 2}\sqrt{V_2+V_1+1/4}, s={1\over 2}\sqrt{V_2-(V_1+1/4)}, t={1 \over 2} \sqrt{1/4+V_1-V_2}.
\end{equation}
Then following [4,5], we can write the transmission amplitude for (1) as
\begin{eqnarray}
t(k,V_1,V_2)={\Gamma[1/2-r-i(s+k)] \Gamma[1/2+r+i(s-k)]\Gamma[1/2+r-i(s+k)]\Gamma[1/2-r+i(s-k)] \over \Gamma[-ik] \Gamma[1+ik]\Gamma^2[1/2-ik]},\\ \nonumber 
r(k,V_1,V_2)=f(k,V_1,V_2) t(k,V_1,V_2), f(k,V_1,V_2) \ne f(k,V_1,-V_2).
\end{eqnarray}
The factor $f(k,V_1,V_2)$ which is unimportant here may be seen in [6].
Earlier for non-symmetric complex potentials handedness of reflectivity has been proved [9].
For complex PT-symmetric potential (1), $r(k,V_1,V_2) \ne r(k,V_1,-V_2)$ follows
consequently [6,9].
\par The property that $\Gamma(-N)= \infty$ where N is non-negative integer helps
in studying the poles of the transmission amplitude $t(k,V_1,V_2)$ (3). The poles
of four Gamma functions in (3) are $ik_n=\pm [n+1/2-(r \pm is)], n=0,1,2,...$ 
and we get a discrete spectrum of complex conjugate pairs [3]
\begin{equation}
E_n=-[n+1/2-(r \pm is)]^2, 
\end{equation}
as $s$ is real. When $V_2 < V_1+1/4$ and $s$ is purely imaginary, we recover [3]
two branches of real discrete spectrum
\begin{equation}
E_{n^+}=-[n^{+}+1/2-(r+t)]^2, E_{n^-}=-[n+1/2-(r-t)]^2, n^{\pm} = 0,1,2,3...m^{\pm}
\end{equation}  
Here $m^{\pm}=[r\pm t], [.]$ denoting the integer part. 
Notice that in these eigenvalues (4,5) the real part is negative.
When $V_1 < 0$, $t$ (2,5) becomes non-real and there are no real bound states.
More importantly in this case the real part of (1) becomes a barrier [6]. 
However, in what follows now $V_1$ could be positive or negative. 
\par Now we find a very interesting scope for the poles of (3) at positive
discrete energies. We set 
\begin{equation}
1/2-r=-n ~ \mbox{or} ~ 1/2+r=-n, n>1
\end{equation}
We get a condition on the potential parameter as
\begin{equation}
V_1+|V_2|=4n^2+4n+{3 \over 4}, n=1,2,3... .
\end{equation}
Further, we get $k=\pm s$ or equivalently
\begin{equation}
E_*={1 \over 4}[|V_2|-(1/4+V_1)],
\end{equation}
which is a single energy. The presence of $|.|$ indicates the commonness of 
these results (7,8), even if the sign of $V_2$ is changed. Changing the sign of
$V_2$ in (1) is equivalent to changing the direction of incidence of the particle
at the potential. In doing so, as said earlier, only the reflection amplitudes 
will change. Also note that for the positivity of $E_*$ and hence for the existence
of the  spectral singularity, $|V_2|$ needs to be larger than $|V_1|+
{\mbox{sgn}(V_1)\over 4}$, meaning the imaginary part of the potential (1) should be 
stronger than the real part. However, for the binding potentials such as 
$V(x)=x^2-g(ix)^\nu$ whose both real and imaginary parts diverge asymptotically, 
the concept of stronger real/imaginary part may not make a sense.
\par
So we conclude that whenever the complex PT-symmetric scattering potential (1) has
its parameters satisfying the condition (7), there will occur a zero-width
single resonance at real energy $E=E_*$ (8). We conjecture that for complex 
PT-symmetric scattering potentials $(s.~t., ~ V(\pm \infty)=0$) the imaginary
part of the potential ought to be necessarily stronger than that of the 
real part. It is instructive to note here that whether or not the real part of the
complex PT-Symmetric potential is a well or a barrier the parameter dependent 
spectral singularity can occur.
Curiously enough like the
model of [8] here too we get a single resonant energy (8) when the potential (1)
is fixed as per the condition (7). Therefore, further, it is desirable to 
investigate whether a complex PT-symmetric potential can support at most one (or more)
spectral singularity(ies).  
\section*{References}
\begin{enumerate}
\item C. M. Bender and S. Boettcher 1998, Phys. Rev. Lett. {\bf 80}, 5243.
\item Z. H. Musslimani, K. G. Makris, R. El-Ganainy, and D.N. Christodoulides 2008,
Phys. Rev. Lett. {\bf 100}, 103904; K. G. Makris, R. El-Ganainy, D.N. Christodoulides,
and Z. H. Musslimani 2008, Phys. Rev. Lett. {\bf 100}, 103904.
\item Z. Ahmed 2001, Phys. Lett. A {\bf 282} 343; {\bf 287} 295.
\item A. Khare and U. P. Sukhatme 1988, J. Phys. A: Math. Gen. {\bf 21} L501.
\item G. Levai, F. Cannata and A. Ventura 2001, J. Phys. A: Math. Gen. {\bf 34} 839.
\item Z. Ahmed 2004, Phys. Lett. A {\bf 324} 152. 
\item A. Mostafazadeh and H. Mehr-Dehnavi 2009, J. Phys. A {\bf 42} 125303. 
\item A. Mostafazadeh 2009, Phys. Rev. Lett. {\bf 102} 220402.
\item Z. Ahmed 2001, Phys. Rev. A 64, 042716.
\end{enumerate} 
\end{document}